\documentstyle[preprint,tighten,aps]{revtex}
\begin{document}
\preprint{\vbox{ \null\hfill INFNCA-TH9602 \\
\null\hfill hep-ph/9601386}}
%
%
\draft
\title{$\protect\bbox{^1\!P_1}$ charmonium state decay
into $\protect\bbox{p\,\bar p}$ in QCD models \\
including constituent quark mass corrections}

\author{Francesco Murgia}

\address{Dipartimento di Scienze Fisiche, Universit\`{a}
di Cagliari, \\
via Ospedale 72, I-09124 Cagliari, Italy \\
and \\
Istituto Nazionale di Fisica Nucleare, Sezione di Cagliari, \\
via Ada Negri 18, I-09127 Cagliari, Italy}

\date{June 1996}

\maketitle

\begin{abstract}

Stimulated by the experimental observation, made by the E760
Collaboration at Fermilab, of the $^1\!P_1$
state of charmonium resonantly formed in $p\,\bar p$
annihilation, we perform a calculation
of the decay width for the $^1\!P_1\to p\,\bar p$ process.
To this end, we employ a phenomenological model
which adds constituent quark mass corrections to the usual massless
QCD models for exclusive processes. For massless models, in fact,
the process under consideration is forbidden by the so-called helicity
selection rules, while it is allowed in our extended
model. We find $\Gamma(^1\!P_1\to p\,\bar p)$ to be in the
range $1-10$ eV. We also compare our results with previous, indirect
estimates, based on QCD multipole expansion models.

\end{abstract}


\narrowtext

\section{Introduction}
\label{intro}

Recently the E760 Collaboration at Fermilab \cite{e760}
have reported the first possible observation of the $^1\!P_1$
bound state of charmonium (this resonance,
which possibly needs further confirmation,
has been indicated as $h_c(1P)$
in the last Review of Particle Properties \cite{pdg94}).
The experimental study of the $^1\!P_1$ charmonium state is very
interesting for testing QCD theoretical models.
The position of its mass with respect to the center of
gravity of the $\chi_c(^3P_J)$ states provides a way to
test different potential models for the $c\,\bar c$ bound
state. Furthermore, the branching ratios for the $^1\!P_1$
hadronic exclusive decays relate to the validity
of the so-called QCD helicity selection rules (HSR)
\cite{bro89}
and of the QCD multipole expansion models (see, e.g.,
Ref.\cite{kua} and references therein).
In particular, the $^1\!P_1 \to p\,\bar p$ decay is forbidden by the
HSR's in massless QCD models for exclusive hadronic processes
at high transfer momentum \cite{bro89} and, due to time reversal
invariance,
the same should hold for the exclusive formation process
$p\,\bar p \to$ $^1\!P_1$, in contradiction with the results
of Ref.~\cite{e760}, if the identification of the resonance
observed at Fermilab with the $^1\!P_1$ state is correct.

As a matter of fact, while massless QCD models
compare reasonably well with
experimental measurements in several cases, they
are in disagreement with the data for a number of processes,
particularly with those violating the HSR's.
We must bear in mind, however, that these models and their
main qualitative results, like the QCD Counting
Rules \cite{bro89} or the
HSR's, are in principle reliable at very high transfer momentum,
while the available experimental results are all in a range of
energies where their full validity can be at least questionable.

Several higher twist effects can in principle be responsible for the
violation of the HSR's. If implemented in the models,
they could give, at least in some cases,
a way to reconcile theoretical predictions
and experimental measurements.
Unfortunately, a consistent inclusion of all higher
twist effects is far from being a trivial work.
In the last years, however, a few extensions of  massless QCD
models have been proposed, in order to study the possible role
played by some of these higher twist effects.
As an example, the diquark model for baryons is one of these
(see, e.g., Ref.~\cite{diq} for a general overview on this topic).
In previous work \cite{a92,a93,mm} we have instead
proposed a different, phenomenological model
that extends the original massless QCD models
by including constituent quark mass effects
for the light (as compared to
the transfer momentum scale in the process) hadrons produced
in the final state.
In this model the valence quarks of the final hadrons
are given a (constituent) mass, $m_q = x_q m_H$,
where $x_q$ is the light-cone fraction of the hadron momentum
carried by the quark. In the small-intermediate
$Q^2$ region, in fact, one may think that the valence, constituent
quarks (the current quarks surrounded by their cloud of
$q\,\bar q $ pairs and gluons) still act as a single, effective particle.
As shown by Weinberg \cite{wei}, these constituent quarks
can indeed be  treated as bare Dirac particles, with the
same coupling as for current quarks in the standard 
model QCD Lagrangian.

This model introduces spin-flip contributions
in the elementary amplitudes
for the hard scattering process, and
as a consequence violations of the HSR's are in principle allowed.
This is possibly the simplest extension of the ordinary massless models,
and can be easily compared with them. It is assumed to be useful in the
intermediate range of momentum transfer $Q^2$ where,
like in charmonium decays,
the use of QCD models is reasonable but at the same time
higher twist effects, of the order $m_H^2/Q^2$
($m_H$ being a hadronic mass scale), may still play
an important role. As the squared momentum transfer $Q^2$ increases,
the predictions of the model smoothly tend to those of the
usual massless models.

This model has been previously applied to
several interesting processes \cite{a92,a93,mm},
in order to test the role of higher twist effects at the
scale of momentum transfer presently accessible by experiments.
For example, it predicts a decay width for the HSR-violating process
$\chi_{c0}\to p\bar p$ comparable with those relative to the
HSR-allowed decays $\chi_{c1,c2}\to p\bar p$ \cite{a92}
(however, only a huge experimental upper limit is available
for $\Gamma(\chi_{c0}\to p\bar p)$, so that this prediction cannot be
severely tested at present);
it also reproduces satisfactorily the experimental angular distribution
of the $p\bar p$ pair in the decay $J/\psi\to p\bar p$, where the role
of HSR-violating contributions is essential (see Ref.~\cite{mm} and
references therein).

However, we must stress that other nonperturbative effects
than the higher twist effects implemented in the model proposed here
can also be relevant for charmonium decays.
In Ref. \cite{a92}, e.g., it has been shown that constituent mass
corrections to the usual perturbative QCD models are not
sufficient to explain the large experimental width for the
HSR-violating decay $\eta_c\to p\bar p$. In fact, the $\eta_c$
state seems to be very peculiar, and in order to explain its,
somehow surprisingly, large decay widths into $p\bar p$ or
vector meson pairs, nonperturbative explanations,
like strong gluonic components \cite{gen} or instanton-induced
contributions \cite{for}, have been proposed. 
The so-called $\rho\pi$ puzzle \cite{bro89} and the radiative decays
of the $J/\psi$ \cite{kua2} offer other examples of processes where
nonperturbative effects can play a substantial role.

It is well possible that different higher twist or nonperturbative effects,
like those quoted above, affect charmonium decays, each of them
being perhaps more relevant for some specific process and less
for others. A detailed comparison of different, theoretical models with
all the available experimental results is then required to improve our
present understanding of charmonium decays.

Following this line of research, in this paper we apply
the model of Ref. \cite{a92,a93,mm} to the calculation of the
decay width for the charmonium decay process $^1\!P_1 \to
p\,\bar p$, whose interest has been briefly recalled before.

The plan of the paper is the following: in Sect. II we
present the derivation of the decay width
for the process $^1\!P_1 \to p\,\bar p$ in the framework of
QCD models including constituent quark mass effects.
Since this can also be useful as a general example
of how these kind of calculations are performed,
we will give a detailed presentation of all the crucial
steps of the calculation.
In Sect. III we present the results in a form which possibly
exploit the model at its best, minimizing whenever
possible model ambiguities.
Finally, in Sect. IV we give some final comments and remarks.

\section{Evaluation of the decay width for the process
\protect$\bbox{^1\!P_1 \to \lowercase{p\,\bar p}}$}
\label{theo}

The general expression of the differential decay width for the process
$^1\!P_1 \to p\,\bar p$ is the following:

\begin{eqnarray}
\frac{d\Gamma(^1\!P_1\to p\bar p)}{d\phi d(\cos\theta)} & = &
\frac{1}{8(2\pi)^5}(1-4\epsilon_1^2)^{1/2} \!\!\!\!
\sum_{M,M',\lambda,\lambda'} \rho_{MM'}(^1\!P_1) \nonumber \\
& \; \times & A_{\lambda,\lambda';M}(\theta,\phi)
A^*_{\lambda,\lambda';M'}(\theta,\phi) \ ,
\label{dg2}
\end{eqnarray}

\noindent where $\theta$ and $\phi$ are respectively the polar and
azimuthal angles defining the outgoing direction of the final
proton, in the $^1\!P_1$ rest frame;
$\epsilon_1$ is the ratio between the proton mass,
$m_p$, and the $^1\!P_1$ mass, $M_1$; $\rho(^1\!P_1)$ is the
spin density matrix of the $^1\!P_1$; finally,
$A_{\lambda,\lambda';M}(\theta,\phi)$ is the rest frame
helicity amplitude for the decay of the $^1\!P_1$ bound state,
with $z$ component $M$ of its total angular momentum, $J=1$,
into a proton and an antiproton whose
helicities are respectively $\lambda$ and $\lambda'$.

It is known from first principles \cite{bour} that the amplitudes
$A_{\lambda\lambda';M}(\theta,\phi)$ have
the following general structure:

\begin{equation}
A_{\lambda\lambda';M}(\theta,\phi) =
\tilde A_{\lambda\lambda'}\,
d^{1}_{M,\lambda-\lambda'}(\theta)
\,\exp(iM\phi) \; ,
\label{agen}
\end{equation}

\noindent where the ``reduced'' amplitude
$\tilde A_{\lambda\lambda'}$
is independent of $M$ and the angular
variables and the $d^{\ J}(\theta)$ are the usual
spin rotation matrices.

Inserting Eq.~(\ref{agen}) into Eq.~(\ref{dg2}) and integrating
over $d\phi$ we get, since $\int_0^{2\pi}d\phi \exp[i(M-M')\phi]
= 2\pi\delta(M-M')$,

\begin{eqnarray}
\frac{d\Gamma(^1\!P_1\to p\bar p)}{d(\cos\theta)} & = &
\frac{1}{8(2\pi)^4}(1-4\epsilon_1^2)^{1/2}
\sum_{\lambda,\lambda'} |\tilde A_{\lambda\lambda'}|^2
\nonumber \\ & \;\; \times &
\sum_M \rho_{MM}(^1\!P_1) \left[ d^1_{M,\lambda-\lambda'}
(\theta) \right]^2 \; .
\label{dg1}
\end{eqnarray}

Integrating now over $d(\cos\theta)$ and using the facts that
$\int_0^\pi d(\cos\theta)
\left[d^J_{\mu\mu'}(\theta) \right]^2 = 2/(2J+1)$, whatever $\mu$
and $\mu'$ are and that, by parity invariance,
$|\tilde A_{--}|^2 = |\tilde A_{++}|^2$ and
$|\tilde A_{-+}|^2 = |\tilde A_{+-}|^2$, we finally find:

\begin{equation}
\Gamma(^1\!P_1\to p\bar p) = \frac{1}{96\pi^4}(1-4\epsilon_1^2)^{1/2}
\left[ |\tilde A_{++}|^2 + |\tilde A_{+-}|^2 \right] \; .
\label{g}
\end{equation}

Before proceeding with calculations, let us remind
why massless QCD models
predict a vanishing decay width for the process under
consideration. First of all,
amplitudes with the proton and antiproton
having opposite helicities, $\tilde A_{\lambda,-\lambda}$,
must vanish: in fact, the initial state of the process, the
$^1\!P_1$, is a $J^{PC} = 1^{+-}$ bound state;
on the other hand, the outgoing $p\,\bar p$ pair is,
for $\tilde A_{\lambda,-\lambda}$, in a $S=1$ state.
Then, in order for the pair to have 
charge conjugation $C=(-1)^{L+S}=-1$, its
relative angular momentum $L$
should be even, contradicting the requirement imposed by
parity conservation, $P = (-1)^{L+1}$, that $L$ should be odd
\cite{note}.
The crucial point is now that in massless QCD models 
helicity amplitudes with the proton and the antiproton
having the same helicity, $\tilde A_{\lambda,\lambda}$,
are also vanishing, due to the HSR's, which forbid spin-flips
in the elementary quark-gluon vertices.
Then there is not any term contributing
to the total decay width, Eq.~(\ref{g}).
Notice that HSR's are valid to all orders in the perturbative
expansion in the strong coupling constant, and as such overcoming
their restrictions is not a matter of going to higher order
in this expansion. On the contrary, every higher twist contribution
violating the HSR's would in principle give a non vanishing contribution
to $\tilde A_{\lambda,\lambda}$, of the order $m_H/M_1$,
where $m_H$ is a typical hadronic scale.

We now proceed considering in more details the helicity amplitudes
$A_{\lambda\lambda';M}(\theta,\phi)$ which enter the
calculation of the $\Gamma(^1\!P_1 \to p\,\bar p)$ decay width.
Let us define: {\it i)} $A^{L\,S\,J}_{\lambda,\lambda';M}(\theta,\phi)$
as the helicity amplitude for the decay of a $c\,\bar c$ bound state
with quantum numbers $L, S, J, M$, into a $p\,\bar p$ pair
with helicities $\lambda$, $\lambda'$ respectively; {\it ii)}
$M_{\lambda\lambda';\lambda_c\lambda_{\bar c}}(\bbox{k};\theta,\phi)$
as the helicity amplitude for the decay of two free $c$, $\bar c$
quarks with relative momentum $\bbox{k}$ and helicities $\lambda_c$,
$\lambda_{\bar c}$ into the same final state as before.
Then, in the well-known nonrelativistic approximation,
$A^{L\,S\,J}_{\lambda,\lambda';M}$ can be obtained by integrating
the amplitude $M_{\lambda\lambda';\lambda_c\lambda_{\bar c}}$
over the proper $c\,\bar c$ bound state, nonrelativistic
wave function:

\begin{eqnarray}
\lefteqn{ A^{L\,S\,J}_{\lambda,\lambda';M}(\theta,\phi) =
\sum_{\lambda_c\lambda_{\bar c}} \left(
{2L+1 \over 4\pi} \right)^{1/2}
C^{{1\over2}\;\;{1\over2}\;\;S}_{\lambda_c,\,-\lambda_{\bar c}
\:\lambda}\,C^{L\, S\, J}_{0\:\lambda\:\lambda} } \nonumber \\
& & \times  \int d^3\bbox{k} \,
M_{\lambda\lambda';\lambda_c\lambda_{\bar c}}(\bbox{k};\theta,\phi)
D^{J\,*}_{M\lambda}(\beta,\alpha,0)
\,\psi_C(k) \; ,
\label{anr}
\end{eqnarray}

\noindent where the $C$'s are the Clebsh-Gordan coefficients
required to give the right combination of quantum numbers; 
$\lambda = \lambda_c - \lambda_{\bar c}$;
$\bbox{k} = (k,\alpha,\beta)$ is
the relative momentum between the $c$ and $\bar c $ quarks
and finally $\psi_c(k)$ is the (momentum-space) charmonium
wave function. In particular, for $L=1$ bound states, and taking
a full nonrelativistic approximation (that is, $\bbox{k} \to 0$),

\begin{equation}
\psi_{L=1}(k) = i3\sqrt{2\pi}|R'_1(0)|\frac{1}{k^3}\delta(k) \; ,
\label{psik}
\end{equation}

\noindent where $|R'_1(0)|$ is the absolute value of
the first derivative of the bound state radial
wave function at the origin.

Using Eqs.~(\ref{agen},\ref{anr},\ref{psik}), and
defining $u=\cos\alpha$ we get:

\begin{eqnarray}
& & A_{++;0}(\theta=\phi=0) = \tilde A_{++} = \nonumber \\
& \;\; & i\frac{3^{3/2}}{2}|R'_1(0)|
\lim_{\bbox{k}\to 0}
\frac{1}{k}\int_0^{2\pi}\!\!\!\!\!d\beta\int_{-1}^{+1}
\!\!\!du\ u \nonumber \\
& \;\; \times & \left[ M_{++;++}(\bbox{k};\theta\!=\!\phi\!=\!0)-
M_{++;--}(\bbox{k};\theta\!=\!\phi\!=\!0) \right]\; ,
\label{app}
\end{eqnarray}

\noindent and

\begin{eqnarray}
& & A_{+-;1}(\theta=\phi=0) = \tilde A_{+-} = \nonumber \\
& \;\; & - i\left(\frac{3}{2}\right)^{3/2}\!\!\!|R'_1(0)|
\lim_{\bbox{k}\to 0}
\frac{1}{k}\int_0^{2\pi}\!\!\!\!\!d\beta \ e^{i\beta}\int_{-1}^{+1} 
\!\!\!\!du\ \!\!\left(1-u^2\right)^{1/2} \nonumber \\
& \;\; \times & \left[ M_{++;++}(\bbox{k};\theta\!=\!\phi\!=\!0)-
M_{++;--}(\bbox{k};\theta\!=\!\phi\!=\!0) \right]\; .
\label{apm}
\end{eqnarray}

Notice that from now on we restrict our calculations to the
simple kinematical configuration $\theta = \phi = 0$,
since the angular dependence of the helicity amplitudes
$A_{\lambda,\lambda';M}$ can be deduced from Eq.~(\ref{agen}).

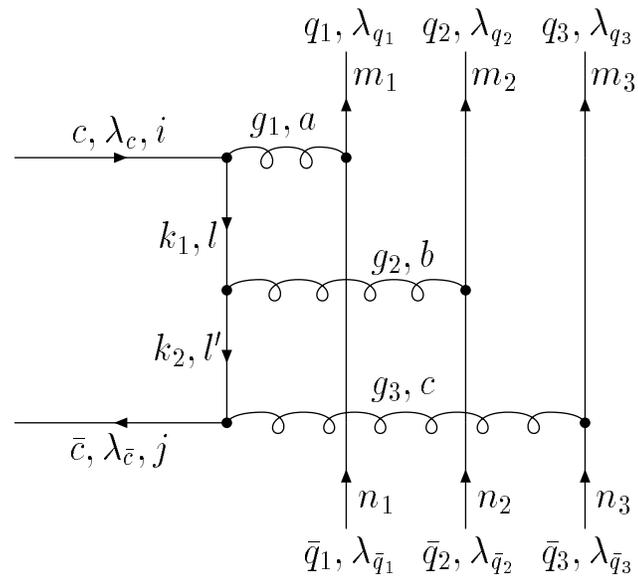
\begin{figure}
\caption{ The Feynman diagram which, to lowest order
in $\alpha_s$, describes the elementary process $Q\bar Q \to
q_1q_2q_3\bar q_1\bar q_2\bar q_3$, for a quarkonium state with
charge conjugation $C=-1$.
In the $Q\bar Q$ center-of-mass
frame, $c^\mu = (E, \bbox{k}/2)$ and
$\bar c^\mu = (E,-\bbox{k}/2)$, where
$\bbox{k}$ is the relative momentum between
the $c$ and $\bar c$ quarks;
$q_{i}=x_{i}p$ and $\bar q_i=y_i\bar p$ ($i=1,2,3$),
with $p^\mu=(E,\bbox{q})$,
$\bar p^\mu=(E,-\bbox{q})$ and
$\bbox{q}=
(q\sin\theta\cos\phi,
q\sin\theta\sin\phi, q\cos\theta)$.
$a,b,c$, $i,j,l,l',m_{1,2,3}$, $n_{1,2,3}$ are color indices;
the $\lambda$'s label helicities.  }
\label{qcd}
\end{figure}

In order to proceed further and evaluate the helicity amplitudes
$M_{\lambda\lambda';\lambda_c\lambda_{\bar c}}(\bbox{k};\theta,\phi)$
we resort to the previously mentioned QCD models \cite{bro89}.
According to them and to factorization theorems,
the amplitudes $M$ are given, under
particular conditions that are reasonably applicable to
charmonium decays, by a convolution between a hard,
perturbative contribution, describing the interaction
between the $c\,\bar c$ pair and the valence quarks of the
final baryons, and a soft, non-perturbative contribution,
which accounts for the hadronization of the valence quarks
into the final hadrons.
The latter contribution is described, for each final hadron,
by the respective distribution amplitude (DA), which
is in principle independent of the
particular process under consideration. As is well
known, the exact form of the DA's is essential for the
quantitative results of the model.
Then we have:

\begin{eqnarray}
& & M_{\lambda\lambda';\lambda_c\lambda_{\bar c}}
(\bbox{k},\theta,\phi) =
\int\ [d\tilde x][d\tilde y]
\psi_{p,\lambda}(\tilde x) \psi_{\bar p,\lambda'}(\tilde y)
\nonumber \\
& \;\;\;\times & T_{\lambda_{q_1}\lambda_{q_2}\lambda_{q_3},
\lambda_{\bar q_1}\lambda_{\bar q_2}\lambda_{\bar q_3};
\lambda_c\lambda_{\bar c}}(\tilde x,\tilde y;\bbox{k},\theta,\phi) \; ,
\label{m}
\end{eqnarray}

\noindent where $\int [d\tilde z]$ stays for $\int_0^1 dz_1dz_2dz_3
\delta(1-z_1-z_2-z_3)$ and we used $\tilde z$
as a shorthand for $(z_1,z_2,z_3)$;
$x_i(y_i)$ represents the (light-cone) fraction of the proton
(antiproton) four-momentum carried by the $i$-th quark (antiquark);
the $\psi(\tilde z)$ are the baryon wave functions, containing
a color, spin-flavor and dynamical part, the above mentioned DA.
We will discuss in more details these wave functions in the following
(see Eq.~(\ref{wfp}) and following comments).
The helicity amplitude $T$ describes the hard elementary scattering.
To lowest order in the strong coupling constant there is only one
Feynman graph contributing (apart from a set of similar terms
obtained by permutations of the final fermionic lines, which
can be accounted for by a proper choice of the hadron wave function).
It is shown in Fig.~\ref{qcd}, where the kinematical notation
is also defined.

The hard scattering amplitude $T$ has the following expression
(we use $\{\lambda\}$ as a shorthand notation for the set
of quark and antiquark helicities,
$\lambda_{q_1}\lambda_{q_2}\lambda_{q_3},
\lambda_{\bar q_1}\lambda_{\bar q_2}\lambda_{\bar q_3}$):

\begin{eqnarray}
T_{\{\lambda\};\lambda_c\lambda_{\bar c}} & = &
- c_F (4\pi\alpha_s)^3\frac{1}{g_1^2g_2^2g_3^2}
\frac{1}{(k_1^2-m_c^2)(k_2^2-m_c^2)} \nonumber \\
& \;\;\;\times & R_{(q)\mu\nu\rho}(\{\lambda\})
R_{(c)}^{\mu\nu\rho}(\lambda_c,\lambda_{\bar c}) \; ,
\label{t}
\end{eqnarray}

\noindent where $c_{_F}$ is the color factor which, once the convolution
with the final hadron wave functions is made, takes the value
$c_{_F} = 5/(18\sqrt{3})$;

\begin{equation}
g_i^2 = M^2_1[x_iy_i+(x_i-y_i)^2\epsilon_1^2]
\label{gi2}
\end{equation}

\noindent is the squared four-momentum
of the $i$-th virtual gluon, $i=1,2,3$;
$k_1= c-g_1$ and $k_2= g_3-\bar c$ are
the four-momenta of the two virtual charm quarks
(see Fig.~\ref{qcd}); $m_c$ is the charm quark mass (we assume,
consistently with the adopted nonrelativistic approximation for the
bound state, $m_c \simeq M_1/2$); moreover,

\begin{equation}
 (k_1^2-m_c^2)(k_2^2-m_c^2) =
 \frac{M_1^4}{4}(A + B\hat k u + C\hat k^2 u^2) \; ,
\label{k}
\end{equation}

\noindent where $\hat k = k/M_1$, and we have defined

\begin{eqnarray}
A & \!\!\!\!\!\!\!\!=\!\!\!\!\!\!\!\! &
    \prod_{i=1,3} \Big[ 2x_iy_i-x_i-y_i
    +2(x_i-y_i)^2\epsilon_1^2\Big]\; ; \nonumber \\
B & \!\!\!\!\!\!\!\!=\!\!\!\!\!\!\!\! &
    \left(1-4\epsilon_1^2\right)^{1/2}\left\{
    (x_1-y_1)\Big[2x_3y_3-x_3-y_3 \right. \nonumber \\
  & \qquad\qquad + & \left.  2(x_3-y_3)^2\epsilon_1^2\Big]
    - (1 \longleftrightarrow 3) \right\} \; ; \nonumber \\
C & \!\!\!\!\!\!\!\!=\!\!\!\!\!\!\!\! &
     -\left(1-4\epsilon_1^2\right)(x_1-y_1)(x_3-y_3) \; ;
\label{abc}
\end{eqnarray}

\noindent finally

\begin{equation}
R_{(q)}^{\mu_1\mu_2\mu_3}(\{\lambda\}) = \prod_{i=1}^3
\bar u(q_i,\lambda_{q_i})\gamma^{\mu_i}
v(\bar q_i,\lambda_{\bar q_i}) \; ,
\label{rq}
\end{equation}

\noindent and

\begin{eqnarray}
R_{(c)}^{\mu\nu\rho}(\lambda_c,\lambda_{\bar c}) & = &
\bar v(\bar c,\lambda_{\bar c})\gamma^\mu(k_2\!\!\!\!\!\!
/\ \ +m_c)\nonumber \\
& \;\;\times & \gamma^\nu(k_1\!\!\!\!\!\!/\ \ +m_c)\gamma^\rho
u(c,\lambda_c) \; .
\label{rc}
\end{eqnarray}

Notice that Eq.~(\ref{g}) implies that the helicity spinors for
the $c$, $\bar c$ quarks are normalized to one, while those
for the final quarks are normalized to $2m$.

Inserting Eqs.~(\ref{m}),(\ref{t}) into
Eqs.~(\ref{app}),(\ref{apm}) we find:

\begin{eqnarray}
\tilde A_{++} \! & = & \! - i \frac{3^{3/2}}{2}\,|R'_1(0)|
\,c_F\,(4\pi)^3\!\!\int_0^1 \!\!\!\![d\tilde x]\,[d\tilde y]\,
\psi_{p,+}(\tilde x) \nonumber \\ & \;\;\;\;\times &
\psi_{\bar p,+}(\tilde y)
\left[ \alpha_s^3\frac{1}{g_1^2g_2^2g_3^2}
R_{(q)\mu\nu\rho} \right]_{k=0}
\hat F_{(0)}^{\mu\nu\rho} \; ;
\label{appf}
\end{eqnarray}

\begin{eqnarray}
\tilde A_{+-} \! & = & \! i \left(\frac{3}{2}\right)^{3/2}
\!\!\!|R'_1(0)|\,c_F\,(4\pi)^3\!\!\int_0^1
\!\!\!\![d\tilde x]\,[d\tilde y]\,
\psi_{p,+}(\tilde x) \nonumber \\ & \;\;\;\;\times &
\psi_{\bar p,-}(\tilde y)
\left[ \alpha_s^3\frac{1}{g_1^2g_2^2g_3^2}
R_{(q)\mu\nu\rho} \right]_{k=0}
\hat F_{(1)}^{\mu\nu\rho} \; ,
\label{apmf}
\end{eqnarray}

\noindent where we have defined ($n=0,1$)

\begin{eqnarray}
\hat F_{(n)}^{\mu\nu\rho} & = & \frac{4}{M_1^5} \lim_{\hat k\to 0}
\frac{1}{\hat k}\int_{-1}^{+1} \!\!\!d u\,
\frac{u^{1-n}\left(1-u^2\right)^{n/2}}
{A + B\hat k u + C\hat k^2 u^2} \nonumber\\
& \qquad\times & \int_0^{2\pi}\!\!\!\!d\beta\,
\exp(in\beta)\,F^{\mu\nu\rho}\;\;,
\label{fn}
\end{eqnarray}

\noindent and

\begin{equation}
F^{\mu\nu\rho} = R_{(c)}^{\mu\nu\rho}(+,+) -
R_{(c)}^{\mu\nu\rho}(-,-)\; .
\label{f}
\end{equation}

Notice that in Eqs.~(\ref{appf}),(\ref{apmf}) the strong coupling
constant $\alpha_s$ has been left inside the convolution
integral. In fact, strictly speaking, its (running) value depends on
the momentum fractions $x_i$, $y_i$ carried by the quarks
participating to the hard scattering, as will be discussed
in the next section. Furthermore, the color part of the
baryon wave functions $\psi_{p,\bar p}$ is now included
into the color factor $c_F$.

After some length but straightforward algebra we get

\widetext
\begin{eqnarray}
\tilde A_{++} & = & i2^9\sqrt{3}|R'_1(0)|
c_F\pi^4 \frac{1}{M_1^9}\int_0^1[d\tilde x]\,[d\tilde y]\,
\psi_{p,+}(\tilde x)\psi_{\bar p,+}(\tilde y)
\prod_{i=1}^3 \frac{\alpha_s(g_i^2)}
{x_iy_i+(x_i-y_i)^2\epsilon_1^2}\frac{1}{A}
\nonumber \\ & \;\;\;\times &
\Bigg\{\frac{B}{A}\bigg[\Big[x_1y_3+x_3y_1
+2(x_1-y_1)(x_3-y_3)\epsilon_1^2\Big]
\Big[R_{(q)123}-R_{(q)213}+R_{(q)312}-R_{(q)321}\Big]
\nonumber \\ & \;\;\; - &
\Big[x_1x_3+y_1y_3-2(x_1-y_1)(x_3-y_3)\epsilon_1^2\Big]
\Big[R_{(q)132}-R_{(q)231}\Big] \bigg]
\nonumber \\ & \;\;\;+ &
\left(1-4\epsilon_1^2\right)^{1/2}\Big[(x_1-y_1)
(R_{(q)312}-R_{(q)321})-(x_3-y_3)(R_{(q)123}-R_{(q)213})
\Big]\Bigg\}\; ;
\label{appr}
\end{eqnarray}

\begin{eqnarray}
\tilde A_{+-} & = & i2^8\sqrt{6}|R'_1(0)|
c_F\pi^4 (1-4\epsilon_1^2)^{1/2}\frac{1}{M_1^9}
\int_0^1[d\tilde x]\,[d\tilde y]\,
\psi_{p,+}(\tilde x)\psi_{\bar p,-}(\tilde y)
\nonumber \\ & \;\;\;\times &
\prod_{i=1}^3 \frac{\alpha_s(g_i^2)}
{x_iy_i+(x_i-y_i)^2\epsilon_1^2}\frac{1}{A}
\Big\{(x_1-y_1)\left[R_{(q)112}-iR_{(q)221}\right]
\nonumber \\ & \;\;\; + &
(x_2-y_2)\left[R_{(q)121}-iR_{(q)212}\right]+
(x_3-y_3)\left[R_{(q)211}-iR_{(q)122}\right] \Bigr\} \; .
\label{apmr}
\end{eqnarray}

Considering now the explicit expression of the
most general spin-flavor component of the proton
wave function (see, e.g., Ref.~\cite{cer89} and references therein),

\begin{eqnarray}
\psi_{p, \lambda}(\tilde{x}) & = & 2\lambda
\frac{F_N}{4\sqrt{6}}\{\varphi(123)u_{1,\lambda}(1)
u_{2,-\lambda}(2)d_{3,\lambda}(3)
+ \varphi(213)u_{1,-\lambda}(1)u_{2,\lambda}(2)
d_{3,\lambda}(3) \nonumber \\
& - &  2T(123)u_{1,\lambda}(1)u_{2,\lambda}(2)
d_{3,-\lambda}(3)
+ (1 \longleftrightarrow 3) + (2 \longleftrightarrow 3)\ \} \; ,
\label{wfp}
\end{eqnarray}

\noindent where $F_N$ is the so-called nucleon decay constant,
$\varphi(\tilde x)$ is the distribution amplitude,
$2T(123) = \varphi(132)+\varphi(231)$ (in obvious notation,
$\varphi(ijk)=\varphi(z_i,z_j,z_k)$),
we find that, being $\Phi(\tilde x,\tilde y)$ a generic
function of $\tilde x$, $\tilde y$,

\begin{eqnarray}
\int [d\tilde x]\, [d\tilde y]
& &\!\!\!\!\!\!\!\!\psi_{p,+}(\tilde x)
\psi_{\bar p,+}(\tilde y)\Phi(\tilde x,\tilde y)
(R_{(q)123}-R_{(q)213})  =
\frac{F_N^2}{24}M_1^3\epsilon_1\int [d\tilde x]\, [d\tilde y]\,
\Phi(\tilde x,\tilde y) \nonumber \\ & \times & \Big\{
\varphi(123)\varphi(213)-\varphi(213)\varphi(123)+
2\varphi(312)T(132) \nonumber \\ & \;\; - &
2T(132)\varphi(312)-2\varphi(321)T(321)+
2T(321)\varphi(321)\Big\}\; ;
\label{r1}
\end{eqnarray}

\begin{eqnarray}
\int [d\tilde x]\, [d\tilde y]
& &\!\!\!\!\!\!\!\!\psi_{p,+}(\tilde x)
\psi_{\bar p,+}(\tilde y)\Phi(\tilde x,\tilde y)
(R_{(q)312}-R_{(q)321})  =
\frac{F_N^2}{24}M_1^3\epsilon_1\int [d\tilde x]\, [d\tilde y]\,
\Phi(\tilde x,\tilde y) \nonumber \\ & \times & \Big\{
2\varphi(123)T(123)-2T(123)\varphi(123)-
2\varphi(132)T(132) \nonumber \\ & \;\; + &
2T(132)\varphi(132)-\varphi(321)\varphi(231)+
\varphi(231)\varphi(321)\Big\}\; ;
\label{r2}
\end{eqnarray}

\begin{eqnarray}
\int [d\tilde x]\, [d\tilde y]
& &\!\!\!\!\!\!\!\!\psi_{p,+}(\tilde x)
\psi_{\bar p,+}(\tilde y)\Phi(\tilde x,\tilde y)
(R_{(q)132}-R_{(q)231})  =
\frac{F_N^2}{24}M_1^3\epsilon_1\int [d\tilde x]\, [d\tilde y]\,
\Phi(\tilde x,\tilde y) \nonumber \\ & \times & \Bigr\{
2\varphi(213)T(123)-2T(123)\varphi(213)+
\varphi(132)\varphi(312) \nonumber \\ & \;\; - &
\varphi(312)\varphi(132)-2\varphi(231)T(321)+
2T(321)\varphi(231)\Big\}\; ;
\label{r3}
\end{eqnarray}

\begin{eqnarray}
\int [d\tilde x]\, [d\tilde y]
& &\!\!\!\!\!\!\psi_{p,+}(\tilde x)
\psi_{\bar p,-}(\tilde y)\Phi(\tilde x,\tilde y)
\left(R_{(q)112}-iR_{(q)221})\right) = 
-\frac{F_N^2}{48}M_1^3\int [d\tilde x]\, [d\tilde y]\,
\Phi(\tilde x,\tilde y)
\nonumber \\ &\times&
\Big\{ \varphi^2(123)+\varphi^2(213)-4T^2(123)-
\varphi^2(132)+\varphi^2(312)+4T^2(132)
\nonumber \\ & + &
\varphi^2(321)-\varphi^2(231)+4T^2(321) \Big\} \; ;
\label{ri1}
\end{eqnarray}

\begin{eqnarray}
\int [d\tilde x]\, [d\tilde y]
& &\!\!\!\!\!\!\psi_{p,+}(\tilde x)
\psi_{\bar p,-}(\tilde y)\Phi(\tilde x,\tilde y)
\left(R_{(q)121}-iR_{(q)212})\right) = 
-\frac{F_N^2}{48}M_1^3\int [d\tilde x]\, [d\tilde y]\,
\Phi(\tilde x,\tilde y)
\nonumber \\ &\times&
\Big\{ -\varphi^2(123)+\varphi^2(213)+4T^2(123)+
\varphi^2(132)+\varphi^2(312)-4T^2(132)
\nonumber \\ & - &
\varphi^2(321)+\varphi^2(231)+4T^2(321) \Big\}\; ;
\label{ri2}
\end{eqnarray}

\begin{eqnarray}
\int [d\tilde x]\, [d\tilde y]
& &\!\!\!\!\!\!\psi_{p,+}(\tilde x)
\psi_{\bar p,-}(\tilde y)\Phi(\tilde x,\tilde y)
\left(R_{(q)211}-iR_{(q)122})\right) = 
-\frac{F_N^2}{48}M_1^3\int [d\tilde x]\, [d\tilde y]\,
\Phi(\tilde x,\tilde y)
\nonumber \\ &\times&
\Big\{ \varphi^2(123)-\varphi^2(213)+4T^2(123)+
\varphi^2(132)-\varphi^2(312)+4T^2(132)
\nonumber \\ & + &
\varphi^2(321)+\varphi^2(231)-4T^2(321) \Big\}\; .
\label{ri3}
\end{eqnarray}

In the products of the $\varphi$, $T$ proton distribution
amplitudes appearing in the previous equations,
the first term is always intended to be
a function of $\tilde x$, the second of $\tilde y$, while
the number in brackets give the pertinent permutation of
the $z_1,z_2,z_3$ arguments.
It is important to notice, as we will see in a moment,
that in all the terms concerning the
$\tilde A_{++}$ (respectively $\tilde A_{+-}$)
amplitude the contribution to the convolution
integral coming from the proton and antiproton
distribution amplitudes is totally
antisymmetric (symmetric) under the exchange of $\tilde x$
and $\tilde y$.

Using Eqs.~(\ref{appr}),(\ref{r1})-(\ref{r3}), we finally find,
after some simple algebra:

\begin{eqnarray}
\tilde A_{++} & = & i\frac{5\cdot 2^6}{3^3}\pi^4
\epsilon_1 (1-4\epsilon_1^2)^{1/2}
F_N^2 |R_1'(0)|\frac{1}{M_1^6}
\int [d\tilde x]\, [d\tilde y]
\prod_{i=1}^3\frac{\alpha_s(g_i^2)}{x_iy_i+(x_i-y_i)^2
\epsilon_1^2} \nonumber \\
& \;\;\; \times &\frac{1}{2x_1y_1-x_1-y_1+
2(x_1-y_1)^2\epsilon_1^2}\frac{1}{2x_3y_3-x_3-y_3+
2(x_3-y_3)^2\epsilon_1^2} \nonumber \\ & \;\;\; \times &
\Bigg\{\Bigg[\frac{x_1-y_1}{2x_1y_1-x_1-y_1+
2(x_1-y_1)^2\epsilon_1^2}-\frac{x_3-y_3}
{2x_3y_3-x_3-y_3+2(x_3-y_3)^2\epsilon_1^2}\Bigg]
\nonumber \\ & \;\;\; \times &
\Bigg[\bigg[x_1y_3+x_3y_1+2(x_1-y_1)(x_3-y_3)\epsilon_1^2\bigg]
\bigg[\varphi(123)\varphi(213)+2\varphi(312)T(132)
\nonumber \\ & \;\;\; - &
2\varphi(321)T(321) +
2\varphi(123)T(123)-2\varphi(132)T(132)+
\varphi(231)\varphi(321)\bigg]
\nonumber \\ & \;\;\; - &
\bigg[x_1x_3+y_1y_3-2(x_1-y_1)(x_3-y_3)\epsilon_1^2\bigg]
\nonumber \\ & \;\;\; \times &
\bigg[2\varphi(213)T(123)+\varphi(132)\varphi(312)-
2\varphi(231)T(321)\bigg]\Bigg] \nonumber \\ & \;\;\; + &
(x_1-y_1)\bigg[2\varphi(123)T(123)-2\varphi(132)T(132)+
\varphi(231)\varphi(321)\bigg]
\nonumber \\ & \;\;\; - &
(x_3-y_3)\bigg[\varphi(123)\varphi(213)
+2\varphi(312)T(132)-2\varphi(321)T(321)\bigg]\Bigg\}\; .
\label{appfin}
\end{eqnarray}

\narrowtext

Notice that the $\tilde A_{++}$ amplitude is
proportional to the ratio $\epsilon_1 = m_p/M_1$, so that,
as it must be, it vanishes in the massless case, that is, in the
$(m_q^2/Q^2) \sim (m_p^2/Q^2) \to 0$ limit (remember that
in our model $m_q = x_q m_p$, $x_q$ being the light-cone fraction
of the hadron momentum carried by the quark).

It is also easy to verify explicitly that, as was
discussed previously, the amplitude $\tilde A_{+-}$
is vanishing, as it must be from first principles.
In fact, Eqs.~(\ref{apmr}),(\ref{ri1})--(\ref{ri3})
show that the argument of the integral over $d\tilde x$, $d\tilde y$ 
is totally antisymmetric under the exchange of $\tilde x$
and $\tilde y$.

\section{Results}
\label{results}

We are now equipped with all the ingredients required to perform
estimates of the decay width for the process $^1\!P_1 \to p\,\bar p$.

To begin with, let us recall very briefly some aspects of QCD models
(both in the massless and massive cases) which are essential in determining
their quantitative predictions. As previously mentioned,
a very important component of the calculations are the
DA's.
The so-called asymptotic DA's, as deducible
from perturbative QCD at very large values of
the transfer momentum,
are not guaranteed to be reliable in the range of
energies experimentally accessible
today and presumably also in the near future,
due to the their slow, logarithmic evolution with $Q^2$.
Non-perturbative methods,
like Lattice QCD or QCD sum rules, give information on the
first moments of the DA's that lead
to a number of models, widely used for practical
calculations. We will consider several of these
model DA's, in order to get indications on
the spread of our results due to this ingredient, without
giving prominence to one model over the others.

The value of the first derivative of the nonrelativistic,
radial bound state wave function is another parameter to be fixed.
One may use potential models to estimate this quantity or,
alternatively, fix its value phenomenologically. This can
be done by comparing the theoretical prediction with the
experimental measurement for
some other process which is better understood theoretically
and more accessible experimentally.

Sudakov factors can also change sizably the quantitative predictions.
The inclusion of these factors has been advocated in order
to improve the consistency of the theoretical models for
exclusive processes and answer some criticisms connected
with the end-point contributions (see Ref.\cite{ster} and
references therein).

Furthermore, we do not know whether and how the DA's are
modified due to higher twist, mass correction effects.

In order to improve the significance of our results and,
where possible, to minimize their dependence on all
the abovementioned uncertainties, we choose to evaluate,
rather than the absolute value of the
$\Gamma(^1\!P_1 \to p\,\bar p)$ decay width, its ratio with
the decay width for the analogous process, $\chi_{c2} \to
p\,\bar p$. In other words, we write

\begin{equation}
\Gamma(^1\!P_1\to p\,\bar p) = \left[\frac{\Gamma(^1\!P_1\to p\,\bar p)}
{\Gamma(\chi_{c2}\to p\,\bar p)}\right]
\,\Gamma(\chi_{c2}\to p\,\bar p)\; ,
\label{gam}
\end{equation}

\noindent where we use the theoretical prediction for
$\Gamma(\chi_{c2}\to p\,\bar p)$ in the denominator and the
corresponding experimental measurement in the numerator
of the r.h.s. of Eq.~(\ref{gam}).

This way, we can neglect the dependence of the results on
$F_N$ and $|R_1'(0)|$. Notice that $F_N$ and $|R_1'(0)|$
appear in the expression of $\Gamma(^1\!P_1\to p\,\bar p)$
to the 4th and 2nd power respectively, so that even small
variations in their values might give relatively sizable
changes in our observable.
Possible Sudakov factor effects might also
be mitigated in the ratio.
We are then confident that our numerical estimates for the
$\Gamma(^1\!P_1\to p\,\bar p)$ decay width, given in the form
of Eq.~(\ref{gam}), have the
higher precision attainable with present models.

The decay width $\Gamma(\chi_{c2}\to p\,\bar p)$ has been
evaluated previously in massless QCD models (see Ref.
\cite{cer89} and references therein). We have done the
calculation following the same procedure previously described
for the $\Gamma(^1\!P_1\to p\,\bar p)$ case. Our result is
in agreement with that of Ref.\cite{cer89}.
Notice that this calculation accounts only for the
leading term in a ($m_p/M$) power expansion (in that case
the $(m_p/M)^0$ term, since the process is not forbidden by HSR's).
For consistency, we also keep only the leading, ($m_p/M$), term
in the calculation of $\Gamma(^1\!P_1\to p\,\bar p)$.
It is important to notice that, this way, we avoid also
possible problems connected
with modifications of the DA's when mass corrections are
taken into account, since these modifications do not
contribute at leading order.

Using Eq.~(\ref{g}), and Eq.~(\ref{appfin}) at leading order in
$\epsilon_1$ we have:

\begin{eqnarray}
\Gamma(^1\!P_1\to p\,\bar p) &=&\left(\frac{2}{3}\right)^7 5^2
\pi^4\epsilon_1^2 \left(1-4\epsilon_1^2\right)^{1/2}
\nonumber \\
&\qquad\times& F_N^4 |R_1'(0)|^2 \frac{1}{M_1^{12}}I_{1}^2\; ,
\label{ghc}
\end{eqnarray}

\noindent where $I_{1}$ is the integral appearing in
Eq.~(\ref{appfin}), taken in the limit $\epsilon_1 \to 0$.
Notice that we have retained the mass contribution in the
kinematical factor coming from phase space integration,
$\left(1-4\epsilon_1^2\right)^{1/2}$.

The corresponding expression for the $\chi_{c2} \to
p\,\bar p$ decay is \cite{cer89,misp}

\begin{eqnarray}
\Gamma(\chi_{c2}\to p\,\bar p) &=& \frac{2^{11}}
{3^6\cdot 5}\pi^4\left(1-4\epsilon_2^2\right)^{1/2}
\nonumber \\
&\qquad\times&F_N^4|R_1'(0)|^2
\frac{1}{M_2^{12}}I_{2}^2\; ,
\label{gchi2}
\end{eqnarray}

\noindent where $M_2$ is the mass of the $\chi_{c2}$
and $\epsilon_2=m_p/M_2$.
For completeness, we report the expression of the
integral $I_{2}$ (see also Ref.\cite{cer89})

\widetext

\begin{eqnarray}
I_{2} & = & \int [d\tilde x]\, [d\tilde y]
 \,\alpha_s[x_1y_1M^2]
\alpha_s[x_3y_3M^2] \alpha_s[(1-x_1)(1-y_1)M^2]
\nonumber \\ & \qquad \times &
\frac{x_1+y_1}{x_1y_1x_3y_3(1-x_1)(1-y_1)
\left[x_1(1-y_1)+y_1(1-x_1)\right]^2}
\nonumber \\ & \qquad \times &
\Big\{ \varphi^2(231)+\varphi^2(132)+\varphi^2(123)+
\varphi^2(321)+ 4T^2(123)+4T^2(132)\Big\}\; .
\label{ichi2}
\end{eqnarray}

Inserting Eq.s~(\ref{ghc}),(\ref{gchi2}) into Eq.~(\ref{gam})
we find finally:

\begin{equation}
\Gamma(^1\!P_1\to p\,\bar p) = \frac{5^3}{2^{4}\cdot 3}
\epsilon_1^2\left(\frac{1-4\epsilon_1^2}
{1-4\epsilon_2^2}\right)^{1/2}
\left(\frac{M_2}{M_1}\right)^{12}
\left(\frac{I_1}{I_2}\right)^2
\Gamma^{exp}(\chi_{c2}\to p\,\bar p) \; .
\label{ris}
\end{equation}

\narrowtext

In Table \ref{tab1} we present the results obtained using
Eq.~(\ref{ris}) and the available models for the DA's.
Besides the nonrelativistic (NR)
and asymptotic (AS) DA's \cite{bro89},
we have used model DA's inspired by QCD sum rules methods:
the Chernyak, Zhitnitsky (CZ) model and its improved (COZ)
version, due to Chernyak, Ogloblin, and Zhitnitsky (see Ref.
\cite{cer89} and references therein); the King, Sachrajda
(KS) model \cite{kin87}; the Gari, Stefanis (GS) model
\cite{gar87}; and, finally, the Heterotic (HET) model proposed
by Stefanis and Bergmann \cite{ste92}.
A collection of the explicit expressions of these DA's
can be found in Ref.~\cite{mm}.

The expression used for the strong, running coupling constant
inside the convolution integrals $I_{1}$ and
$I_{2}$ is the following:

\begin{equation}
\alpha_s(Q^2) = \frac{12\pi}
 {(11n_c-2n_f)\log\left[(Q^2+4m_g^2)/\Lambda^2\right]} \quad ,
\label{as}
\end{equation}

\noindent where $n_c=3$ is the number of colors; $n_f$ is
the number of active flavors (in this context, $n_f=4$);
$m_g \sim 0.5$ GeV is an effective, dynamical gluon mass
regularizing the $\alpha_s$ behavior at slow $Q^2$ \cite{cor82};
we assume also $\Lambda = 0.2$ GeV/$c$.

The $Q^2$ evolution of the proton DA's, according to perturbative
QCD, has been also taken into account.

Finally, we have used $\Gamma^{exp}(\chi_{c2}\to p\,\bar p)
= (206 \pm 22)$ eV \cite{pdg94,chi2}.

It is evident from Table \ref{tab1} that, in spite of our efforts in order
to limit the dependence of the numerical results on the particular
shape of the model DA, our estimate for $\Gamma(^1\!P_1\to p\,\bar p)$,
even excluding the NR and AS distribution amplitudes,
still cover a range from 1 to 10 eV.
This spread of the results is in part due to the different
behavior of the integrals $I_1$ and $I_2$ when changing the DA's.
Relative changes are more sensitive for $I_1$ than for $I_2$.
This can be easily understood from a qualitative point of view.
In fact, the integral $I_{1}$ is exactly vanishing for
DA's $\varphi(\tilde z)$ that, like the nonrelativistic and
the asymptotic ones, are completely symmetric under permutations
of $z_1, z_2, z_3$.
Moreover, due to the factors $(x_i-y_i)$
which are present in each contribution to $I_{1}$,
$I_1$ itself is very
depressed when $x_i \sim y_i$, as is the case for DA's
that are strongly peaked at $x_i \sim y_i \sim 1/3$.
Then, the magnitude of $I_{1}$
should give a measure of how much the corresponding DA is
far from the simple, intuitive equipartition of the
hadron momentum among its valence constituents.
In principle this could give an interesting way to
discriminate at least the gross features of the proton
DA's. In practice, however, at this stage of the calculations,
these effects are probably masked by other, less controlled,
sources of uncertainty. From this point of view it should surely
be of relevance to perform a full implementation of
mass corrections. We have not calculated them for the
$\chi_{c2} \to p\,\bar p$ process, while for the $^1\!P_1$ case,
direct use of the full expression of $\tilde A_{++}$,
Eq.~(\ref{appfin}), neglecting possible modifications in
the shape of the DA, gives a sizable reduction in the
total width. The worst case then would be if a
full account of mass corrections in the $\chi_{c2}$
case should give an enhancement of the corresponding decay width,
depressing our prediction for the
relative $\Gamma(^1\!P_1\to p\,\bar p)$
decay width, Eq.~(\ref{gam}). We must not forget, however,
that the results presented in Table \ref{tab1}, which are at leading
order in the mass corrections, are on the contrary
free from the ambiguity connected
with possible changes in the shape of the DA's due to higher order
mass corrections, since they do not contribute to leading order.

Let us now compare our predictions for $\Gamma(^1\!P_1\to p\,\bar p)$
with previous estimates.
Assuming the resonance observed
by the E760 Collaboration to be the $^1\!P_1$
state of charmonium, we have the following measurement for the
product of the branching ratios $B(^1\!P_1\to p\,\bar p)
B(^1\!P_1 \to J/\psi\,\pi^0)$ \cite{e760}:

\begin{eqnarray}
(2.3\pm 0.6)\times 10^{-7} & \geq & B(^1\!P_1\to p\,\bar p)
B(^1\!P_1 \to J/\psi\,\pi^0) \nonumber \\
& \geq & (1.7\pm 0.4)\times 10^{-7}\; .
\label{bran}
\end{eqnarray}

The range of values in Eq.~(\ref{bran}) corresponds to the
plausible range of values for the $^1\!P_1$ total decay
width, $\Gamma_T(^1\!P_1)$ \cite{e760}:

\begin{equation}
500 \;\mbox{keV} \leq \Gamma_T(^1\!P_1) \leq 1000 \;\mbox{keV} \; .
\label{gtot}
\end{equation}

By using Eq.~(\ref{gtot}) and the biggest of our estimates from
Table I, we find correspondingly

\begin{equation}
2.1\times 10^{-5} \geq B(^1\!P_1\to p\,\bar p)
\geq 1.0\times 10^{-5}\; .
\label{mybra}
\end{equation}

Unfortunately we are not able to give a reliable prediction for
the decay width of the process $^1\!P_1\to J/\psi\,\pi^0$ in the
framework of our model. Then, we cannot make a full comparison
of our results with the experimental ones of Ref.\cite{e760}.
A complementary situation is found in the case of the QCD multipole
expansion models. In this framework, Kuang et al. \cite{kua} have
predicted $\Gamma(^1\!P_1\to J/\psi\,\pi^0) = 0.3\,(\alpha_{_M}/
\alpha_{_E})$ keV, where $\alpha_{_E}$ and $\alpha_{_M}$
are effective electric and magnetic multipole-expansion
coupling constants. There is some uncertainty on the value
of the ratio $\alpha_{_M}/\alpha_{_E}$,
but theoretical considerations suggest as reasonable range
$1 < \alpha_{_M}/\alpha_{_E} < 3$ \cite{kua}. We thus take as indicative
estimate $\alpha_{_M}/\alpha_{_E} \simeq 2$, which leads to the prediction
$\Gamma(^1\!P_1\to J/\psi\,\pi^0) \simeq 0.6$ keV.
Comparing this result with Eqs.~(\ref{bran}),(\ref{gtot}),
the following estimate for $B(^1\!P_1\to p\,\bar p)$ follows:

\begin{equation}
1.9\times 10^{-4} \leq B(^1\!P_1\to p\,\bar p)
\leq 2.8\times 10^{-4} \; . 
\label{kubra}
\end{equation}

The two predictions of Eqs.~(\ref{mybra}),(\ref{kubra}) differ
approximately of one order of magnitude; they are closer
the smaller is the total decay
width for the $^1\!P_1$ state.
However, both theoretical calculations
are at present order of magnitude estimates, and before
drawing any definite conclusion better calculations
are to be expected, together with improved experimental
results.

Let us finally remark that our prediction for the branching ratio
$B(^1\!P_1 \to p\,\bar p)$, Eq.~(\ref{mybra}),
compares reasonably well
with the experimental branching ratios for the processes
$\chi_{c2,1}\to p\,\bar p$, $B(\chi_{c2} \to p\,\bar p)
= (10.0\pm 1.0)\times 10^{-5}$,
$B(\chi_{c1} \to p\,\bar p) = (8.6\pm 1.2)\times 10^{-5}$
\cite{pdg94,chi2}.
In fact, in QCD models  the $^1\!P_1 \to p\,\bar p$ decay should
be suppressed with respect to the $\chi_{c2,1}\to p\,\bar p$ ones
by a factor $\sim m_H^2/Q^2$, being $m_H\sim 1$ GeV a typical mass
scale for higher twist effects. For $Q^2 \sim M_{1,2}^2$ this should
correspond to a suppression factor of about $10-15$, which should
in turn put the expected value for $B(^1\!P_1 \to p\,\bar p)$ in the
range covered by our prediction.

\section{Conclusions}
In this paper we have estimated the width for the decay process
of the $^1\!P_1$ charmonium state into a proton-antiproton pair,
using QCD models and including constituent
quark mass corrections. In fact, massless QCD models state that
this decay width is vanishing, due to the helicity selection rules.
This is in possible contradiction with the recent experimental
findings of the E760 Collaboration.
The inclusion of higher twist effects, like constituent
quark mass effects, which can still play
a non negligible role at the involved energies, can in
principle improve considerably the situation.
However, a full, consistent treatment of these effects is out
of our present possibilities.
We have then used a phenomenological model which
extends the massless QCD ones.
In order to minimize the dependence of the
quantitative results of the calculation on details
of the model, we have expressed
the $^1\!P_1\to p\,\bar p$ decay width in connection
with that for the $\chi_{c2}\to p\,\bar p$ decay process:
$\Gamma(^1\!P_1\to p\,\bar p) = [\Gamma(^1\!P_1\to p\,\bar p)/
\Gamma(\chi_{c2}\to p\,\bar p)]^{th}\times
\Gamma^{exp}(\chi_{c2}\to p\,\bar p)$.
That is, we evaluate the first term in this expression
in our model, at leading order in the parameter
$\epsilon = m_p/M$, while taking the available experimental
measurement for the second term. This way, not only some 
phenomenological parameters of the model
(the nucleon decay constant $F_N$, the value of the
first derivative of the radial, $L=1$, nonrelativistic
charmonium bound-state wave function at the origin)
cancel out from the result, but some other
effects should be at least minimized.
For example, the possible effect of Sudakov factors
should be reduced. Furthermore, a possible dependence of the
DA's on mass effects is also irrelevant as long as only
the leading term in the $\epsilon = m_p/M$ power expansion
is kept into account. A consistent inclusion of non-leading terms
in the $\epsilon$  power expansion would in fact
require an estimate of mass effects on the
$\Gamma(\chi_{c2}\to p\,\bar p)$ decay width and on the
DA's. As shown in table \ref{tab1}, we estimate the
$\Gamma(^1\!P_1\to p\,\bar p)$ decay width to be of the
order $1-10$ eV, depending on the DA considered.
The corresponding branching ratio depends of course on the
$^1\!P_1$ total decay width, $\Gamma_T(^1\!P_1)$,
which is not known at present.
Using, as suggested by the E760 Collaboration, the range
500 keV $\leq \Gamma_T(^1\!P_1) \leq $ 1000 keV, and the biggest
of our estimates of Table I, we find
$2.1\times 10^{-5}$ $\geq B(^1\!P_1\to p\,\bar p)$
$\geq 1.0\times 10^{-5}$.
This result is sizably lower (about one order of magnitude)
than the corresponding branching ratios
for the $\chi_{c1,c2}$ states, as measured by the E760 Collaboration. 
This is reasonable, since in QCD models
the decay $^1\!P_1\to p\,\bar p$ should be suppressed
by a factor $m_H^2(\sim 1\mbox{GeV}^2)/
M_1^2 \sim 1/12$ with respect to the decays
$\chi_{c1,2} \to p\,\bar p$.
The above result is also sizably lower than that obtained by comparing
the experimental results of Ref.\cite{e760}
with theoretical estimates
of the decay width for the process $^1\!P_1\to J/\psi +
\pi^0$, obtained in the framework of
QCD multipole expansion models.
Both our calculation and that based on QCD multiple
expansion models can probably be affected by
``theoretical'' errors which can modify their
quantitative predictions by a sizable factor.
The two estimates, at least indicatively, can be taken as
representative of two classes of different models:
the first extending the usual perturbative
QCD models by the inclusion of higher twist effects, like mass corrections;
the latter including effectively some different, nonperturbative effects,
like those present, e.g., in the $\eta_c \to p\bar p$ decay process.
As such, their results should at least suggest a definite
(albeit still large) range of possible values for the
$\Gamma(^1\!P_1\to p\,\bar p)$ decay width,
ranging approximately from some eV to some hundreds of eV.

We conclude observing once more that further, more precise, experimental
measurements will certainly help in clarifying our understanding
of the $^1\!P_1\to p\bar p$ decay process and will be very useful to
improve in general our theoretical models for the
calculation of exclusive decay processes in the charmonium family.

\acknowledgments

The author is very grateful to M. Anselmino for many
useful discussions and comments, and to M. Lissia for
a critical reading of the manuscript.

\begin{table}
\caption{ Predictions, as from Eq.~(\ref{ris}), for the decay width
relative to the process $^1\!P_1\to p\,\bar p$, considering several
available models for the nucleon distribution amplitude (see text for
more details).}
\begin{tabular}{cd}
$DA$ & $\Gamma(^1\!P_1\to p\,\bar p)$ (eV) \\
\tableline
NR  &  0.0 \\
AS  &  0.0 \\
CZ  & 10.5 \\
COZ &  2.7 \\
KS  &  1.2 \\
GS  &  1.3 \\
HET &  4.7 \\
\end{tabular}
\label{tab1}
\end{table}

\begin{references}

\bibitem{e760}	E760 Collaboration, T.A.~Armstrong et al.,
		Phys. Rev. Lett. {\bf 69}, 2337 (1992).

\bibitem{pdg94}	Particle Data Group, L.~Montanet et al.,
		Phys. Rev. D {\bf 50}, 1173 (1994).

\bibitem{bro89}	See, e.g., S.J.~Brodsky and G.P.~Lepage,
		in {\it Perturbative Quantum Chromodynamics},
		edited by A.H. Mueller, (World Scientific,
		Singapore, 1989), and references therein.

\bibitem{kua}	Y.-P.~Kuang, S.F.~Tuan, and T.-M.~Yan,
		Phys. Rev. D {\bf 37}, 1210 (1988).

\bibitem{diq}	M.~Anselmino, E.~Predazzi, S.~Ekelin,
		S.~Fredriksson, and D.B.~Lichtenberg,
		Rev. of Mod. Phys. {\bf 65}, 1199 (1993).

\bibitem{a92}	M.~Anselmino, R.~Cancelliere, and F.~Murgia,
                 Phys. Rev. D {\bf 46}, 5049 (1992).

\bibitem{a93}	M.~Anselmino and F.~Murgia,
		Phys. Rev. D {\bf 47}, 3977 (1993);
		{\bf 50}, 2321 (1994);
                preprint INFNCA-TH9520, hep-ph/9510231,
                unpublished.

\bibitem{mm}	F.~Murgia and M.~Melis,
		Phys. Rev. D {\bf 51}, 3487 (1995).

\bibitem{wei}	S.~Weinberg, Phys. Rev. Lett. {\bf 65},
		1181 (1990).

\bibitem{gen}	M.~Anselmino, M.~Genovese, and E.~Kharzeev,
		Phys. Rev. D {\bf 50}, 595 (1994), and references
		therein.

\bibitem{for}	M.~Anselmino and S.~Forte,
		Phys. Lett. B {\bf 323}, 71 (1994).

\bibitem{kua2}	C.-H.~Chang, G.-P.~Chen, Y.-P.~Kuang, and Y.-P. Yi,
		Phys. Rev. D {\bf 42}, 2309 (1990), and references
		therein.

\bibitem{bour}	See, e.g., C.~Bourrely, E.~Leader, and J.~Soffer,
		Phys. Rep. {\bf 59}, 95 (1980).

\bibitem{note}	This is not the case, e.g., for the $J/\psi
		\to p\,\bar p$ decay, because the $J/\psi$ is a
		$J^{PC} = 1^{--}$ state. There is then no mutual exclusion
		between $L$ values required by simultaneous parity
		and charge conjugation conservation, and the
		$\tilde A_{\lambda,-\lambda}$ can be non-vanishing.

\bibitem{cer89}	V.L.~Chernyak, A.A.~Ogloblin, and I.R.~Zhitnitsky,
		Z. Phys. C {\bf 42},  569 (1989);
		{\bf 42}, 583 (1989).

\bibitem{ster}	G.~Sterman and H.-N.~Li,
		Nuc. Phys. B {\bf 381}, 129 (1992).

\bibitem{misp}	Notice that there is a misprint in Eq. (20) of the
		second of Ref. \cite{cer89}, in that the factor
		$f_N/\bar M^2$ should appear to the 4th, rather
		than to the 2nd power.

\bibitem{kin87}	I.D.~King and C.T.~Sachrajda,
		Nucl. Phys. B {\bf 279}, 785 (1987).

\bibitem{gar87}	M.~Gari and N.G.~Stefanis,
		Phys. Lett. B {\bf 175}, 462 (1986).

\bibitem{ste92}	N.G.~Stefanis and M.~Bergmann,
		Phys. Rev. D {\bf 47}, 3685 (1993).

\bibitem{cor82}	J.M.~Cornwall, Phys. Rev. D {\bf 26}, 1453 (1982).

\bibitem{chi2}	E760 Collaboration, T.A.~Armstrong et al.,
		Nuc. Phys. B {\bf 373}, 35 (1992).

\end{references}
\end{document}